\documentclass[journal]{IEEEtran}
\usepackage{cite}
\usepackage{amsmath,amssymb,amsfonts}
\usepackage{graphicx}
\usepackage{booktabs,array,multirow}
\usepackage{algorithmic}
\usepackage{algorithm}
\usepackage{stfloats}
\usepackage{placeins}
\usepackage[table]{xcolor}
\usepackage[hidelinks]{hyperref}
\hypersetup{pdftitle={Guess and Send: World-Model-Guided Semantic Packet Transmission for Robust Video Perception},pdfauthor={Yan Gong et al.},pdfkeywords={semantic communication, world model, video perception, unequal error protection}}
\graphicspath{{figs/}}
\newcommand{\method}{WorldGT}
\newcommand{\E}{\mathbb{E}}
\newcommand{\warp}{\mathcal{W}}
\definecolor{worldred}{RGB}{179,60,53}
\definecolor{revisiongreen}{RGB}{0,125,35}

\definecolor{worldwash}{RGB}{250,239,237}

\begin{document}
\title{Guess and Send: World-Model-Guided Semantic Packet Transmission for Robust Video Perception}

\author{Yan~Gong, Zheng~Chu, Wei~Koong~Chai, Hari~Mohan~Pandey, Jiankang~Zhang, Jingrui~Hou\\Miao Zhang,~\emph{Member, IEEE}, Yuanjian Li~\emph{Member, IEEE}, and Pei Xiao~\emph{Senior Member, IEEE}%
\thanks{Y. Gong, W. K. Chai, H. M. Pandey, and J. Zhang are with the School of Computing and Engineering, Bournemouth University, Poole, U.K. (e-mail: ygong@bournemouth.ac.uk; wchai@bournemouth.ac.uk; hpandey@bournemouth.ac.uk; jzhang3@bournemouth.ac.uk).}%
\thanks{Z. Chu is with the Department of Electrical and Electronic Engineering, University of Nottingham Ningbo China, Ningbo, China (e-mail: andrew.chuzheng7@gmail.com).}%
\thanks{J. Hou is with the School of Information Management, Wuhan University, Wuhan, China (e-mail: houjingrui@whu.edu.cn).}
\thanks{Miao Zhang is with the School of Information Science
and Engineering, Chongqing Jiaotong University, Chongqing 400074, China (e-mail: msczz@foxmail.com).}
\thanks{Yuanjian Li is with the Academy of Artificial Intelligence and Advanced Technology, Xi’an Jiaotong-Liverpool University, Suzhou 215123, China (e-mail: yuanjian.li@xjtlu.edu.cn)}
\thanks{Pei Xiao is with the Institute for Communication Systems (ICS), University of Surrey, Guildford, Surrey GU2 7XH, United Kingdom (e-mail: p.xiao@surrey.ac.uk).}
}

\markboth{IEEE Journal Manuscript}{Gong \MakeLowercase{\textit{et al.}}: World-Model-Guided Semantic Packet Transmission}
\maketitle
\begin{abstract}
Reliable video perception over bandwidth-constrained wireless
links requires transmitting information that the receiver cannot
already recover from its evolving scene representation.
This paper introduces \method{}, a world-model-guided semantic
packet transmission framework that prioritizes information
according to its contribution to the receiver's perception.
A shared feature-space world model predicts the receiver's
scene state through affine motion compensation and learned
residual displacement, while sender-side verification identifies
reusable predictions. An acknowledgment-synchronized receiver
twin estimates task-impact scores through two counterfactual
evaluations, guiding packet selection and unequal error
protection over LDPC-coded wireless channels.
A compact 3-bit feature representation increases spatial
coverage under stringent bandwidth constraints, while
predictive fallback sustains perception during packet outages.
Experiments on KITTI-MOTS (KITTI) and VisDrone2019-MOT (VisDrone) under Rayleigh fading
and AWGN demonstrate substantial gains in low-SNR conditions.
On KITTI at a 4.13\% reference bandwidth budget,
\method{} achieves 47.5\% J\&F at 0~dB Rayleigh SNR,
outperforming H.264 by 11.3 percentage points.
Controlled ablations show that world-model prediction
contributes up to 8.7 points under Rayleigh fading.
Object-detection experiments further demonstrate the
transferability of receiver-relative predictive
transmission across visual perception tasks.
These results establish world-model-guided,
receiver-state-aware transmission as an effective
approach to robust video perception under severe
communication constraints.
\end{abstract}

\section{Introduction}
\IEEEPARstart{R}{emote} video perception supports connected
vehicles, roadside sensing, and unmanned aerial vehicles
(UAVs), yet the wireless link may be least reliable when
timely perception is most valuable. A camera can reduce
its traffic by transmitting features that support downstream
tasks rather than encoded images for reconstruction at the
receiver. Intermediate visual features provide reusable
representations that support multiple downstream perception
tasks. Efficiently maintaining such representations over
unreliable wireless links is therefore important for
distributed visual perception. Semantic and task-oriented
communication formalize the transmission of task-relevant
information~\cite{gunduz2023beyond,qin2021semantic,
kountouris2021semantics,shao2022learning};
deep joint source-channel coding (DeepJSCC) further
enables visual information to be transmitted over noisy
wireless channels~\cite{bourtsoulatze2019deep,
tung2022deepwive,wang2023dvst}.
For a video receiver with memory, however, task relevance
alone does not determine whether new information needs
to be transmitted.

The receiver already holds a partial description of the scene. Some regions remain valid, others can be advanced using motion, and a smaller subset requires correction. A highly confident object at the sender may need no update if the receiver already tracks it accurately. Conversely, a modest local change may deserve strong protection if it alters an object boundary at the receiver. The central question is therefore: \emph{which packets change what this receiver can infer, given what it has already received?} This question couples temporal prediction, packet selection, and channel protection through the receiver's state.

\begin{figure}[!t]
\centering
\includegraphics[width=0.94\columnwidth]{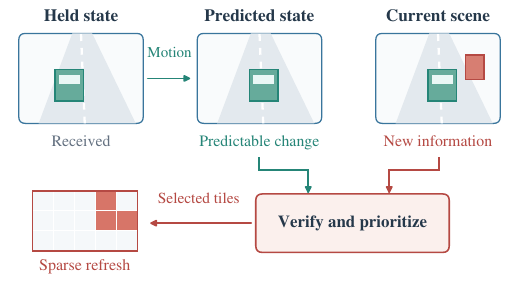}
\caption{Receiver-relative communication principle (conceptual illustration). Motion prediction advances information already held by the receiver. Comparing the prediction with the current observation identifies where a refresh may improve inference. \method{} verifies predictions and prioritizes current feature tiles as the transmitted payload.}
\label{fig:intro}
\end{figure}

Existing approaches address parts of this problem. Feature-change rules~\cite{miskowicz2006send} detect novelty in the transmitted representation but cannot distinguish task-relevant change from predictable camera motion. Confidence-based collaborative perception~\cite{hu2022where2comm,hu2024codefilling} identifies informative regions, although sender confidence does not describe what a remote receiver has accumulated over time. Predictive communication uses temporal models to avoid unnecessary updates~\cite{jiang2025semantic,wang2026world,ercetin2026predictive}. The remaining challenge considered here is to combine receiver-relative packet utility, verified state prediction, and discrete channel protection for dense video perception over a lossy digital link.

We propose \method{} (world-model-guided semantic packet transmission). The sender maintains a synchronized copy of the deterministic receiver, termed a receiver twin. Given codeword acknowledgments, the twin reproduces the receiver's cache and task memory. A shared world model first predicts the cache at the current time. The sender then uses its current observation to verify this prediction tile by tile and evaluates how the receiver's output would change with all packets refreshed versus none refreshed. Aggregating this change spatially yields packet scores for a budget-constrained unequal-error-protection (UEP) allocator with a skip option. Prediction thus provides the reference against which transmission utility is measured.

Fig.~\ref{fig:intro} illustrates the underlying principle: communication should refresh the information that prediction alone cannot supply. \method{} repeats the verification mask with every codeword and uses deterministic fallback prediction during frame outages. Combined with a coarse feature payload, this design increases spatial coverage and maintains temporal continuity under severe losses. The resulting operating objective is strong low signal-to-noise ratio (SNR) perception while remaining competitive as the channel improves.

The contributions are as follows.
\begin{enumerate}
\item We formulate packet transmission for a stateful perception receiver and construct a receiver-relative task-impact proxy from two twin evaluations per frame. The score requires neither current task labels nor a separate receiver evaluation for every packet.
\item We propose a sender-verified feature-space world model that combines signaled affine motion with a factorized spatiotemporal transformer. Its residual displacement and learned gate predict usable cache content from incomplete histories; tile verification and no-arrival rules determine when this content is applied.
\item We integrate prediction and selection with discrete low-density parity-check (LDPC) rate allocation, whole-codeword packing, and a 3-bit payload to prioritize useful spatial updates under a fixed channel budget.
\item We evaluate driving and aerial video under Rayleigh fading and additive white Gaussian noise (AWGN), report gains relative to competing baselines across SNR, and isolate prediction gains using controlled ablations. A detection extension examines the utility of the selected representations beyond the primary segmentation task.
\end{enumerate}

\section{Related Work}\label{sec:related}

\subsection{Semantic and Predictive Communication}
Semantic communication optimizes information delivery
for inference rather than bit-perfect recovery~\cite{gunduz2023beyond,kountouris2021semantics,zhu2026information}.
DeepSC~\cite{xie2021deepsc} and information-bottleneck
methods~\cite{shao2022learning} learn representations
for semantic or task objectives.
DeepJSCC~\cite{bourtsoulatze2019deep} learns direct mappings
to channel symbols, while DeepWiVe~\cite{tung2022deepwive}
exploits temporal dependencies and adaptive bandwidth
allocation for wireless video reconstruction.
Deep video semantic transmission~\cite{wang2023dvst}
further incorporates machine-vision objectives through
learned mappings to analog channel symbols.
However, determining packet utility from the receiver's
accumulated and predicted perception state under previous
transmission losses remains an important consideration.
Age-of-information methods~\cite{kaul2012realtime}
likewise recognize the importance of update timing,
but a dense perception receiver also requires a spatial
decision within each update.

\subsection{Collaborative Perception and Video Analytics}
Feature-sharing systems such as V2VNet~\cite{wang2020v2vnet} and robust task-oriented collaborative perception~\cite{fang2025racp} reduce the cost of exchanging sensor observations. Where2comm~\cite{hu2022where2comm} uses spatial confidence and, in multiple rounds, receiver request maps. CodeFilling~\cite{hu2024codefilling} combines information scores with codebook-based communication. These mechanisms account for spatial demand and complement temporal receiver-state tracking. In edge analytics, Glimpse~\cite{chen2015glimpse} and Reducto~\cite{li2020reducto} filter updates at the camera, whereas DDS~\cite{du2020server} requests selected regions from the server. Estimating the receiver's current demand from its accumulated state offers a complementary perspective on spatial and temporal communication decisions. Confidence-based selection and frame triggering provide representative mechanisms for evaluating these decisions.

\subsection{World Models and Receiver Digital Twins}
World models learn temporal structure for prediction and decision making~\cite{ha2018world,hafner2025mastering,bardes2024vjepa}. In semantic video communication, Jiang \emph{et al.}~\cite{jiang2025semantic} use a video foundation model with depth-based feedback to omit transmissions and repair predicted frames through segmentation-assisted partial updates. Integrating regional feature refresh with digital channel protection remains an important direction for receiver-aware transmission. Wang \emph{et al.}~\cite{wang2026world} evaluate semantic tokens through imagined long-horizon control returns. Their objective is closed-loop control, whereas dense video perception requires immediate task outputs at a fixed receiver. Predictive-state communication~\cite{ercetin2026predictive} studies innovation and reconciliation under delay, and TWIST~\cite{liu2026twist} combines token synchronization with application-aware protection. These approaches motivate shared predictive state and highlight the potential of motion-based feature-cache prediction with sender-side verification. Representing visual dynamics directly in the feature space used by the perception receiver offers a further direction for predictive communication.
However, these studies do not jointly address ACK-synchronized feature-cache prediction, sender-side verification, receiver-relative task-output scoring, and discrete LDPC packet protection for dense video perception. Existing motion-compensation and resource-allocation mechanisms also remain to be integrated within such a receiver-aware communication framework.

\subsection{Error Protection and Resource Allocation}
UEP prioritizes information with unequal distortion consequences~\cite{hamzaoui2005optimized}. Semantic-aware neural video codecs also assign different protection to packet streams~\cite{mortaheb2026semantic}. Lagrangian rate allocation~\cite{shoham1988efficient,ortega1998rate} and the multiple-choice knapsack formulation~\cite{sinha1979multiple} provide established tools for discrete budget allocation. Incorporating receiver-relative task values into these allocation mechanisms could further couple protection decisions with prediction and packet skipping. Similarly, video coding already uses motion compensation and block skip decisions~\cite{wiegand2003overview,sullivan2012overview}. For task-oriented transmission, feature-cache prediction error and task-output changes provide distinct criteria for verification and transmission priority.

\section{System Model and Problem Formulation}\label{sec:system}

\subsection{Overview of WorldGT Framework}
\label{sec:framework_overview}

\begin{figure*}[!tb]
\centering
\includegraphics[width=\textwidth]{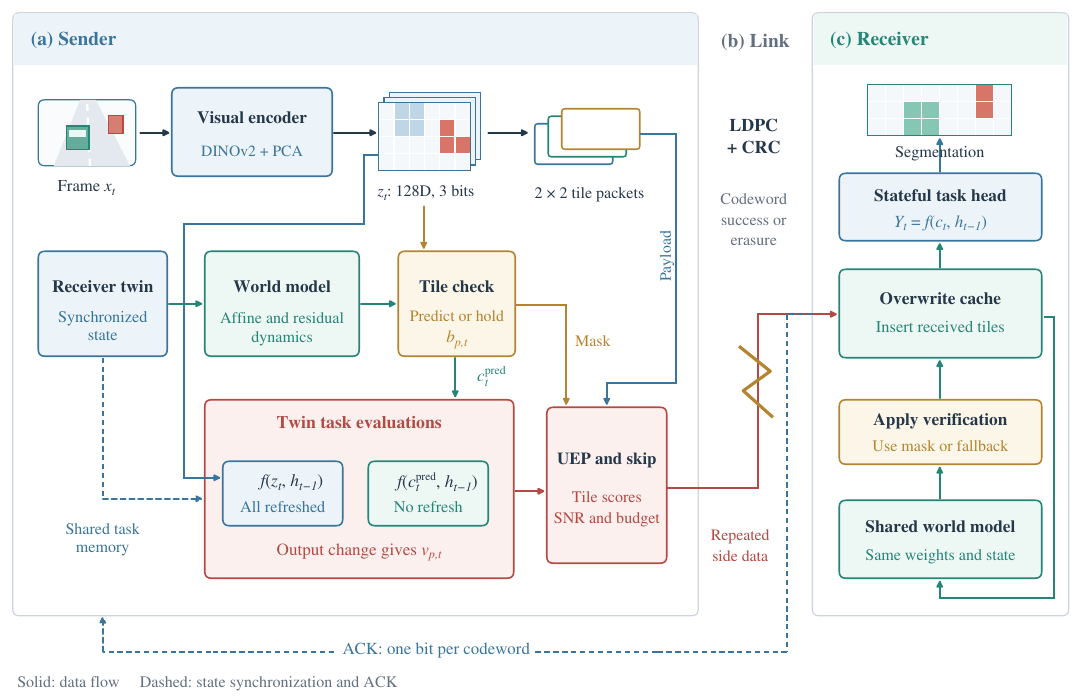}

\caption{\method{} framework. The sender observes the current frame and maintains an acknowledgment (ACK)-synchronized receiver twin. A shared world model predicts the cache; sender verification selects usable tiles. Two task-head evaluations produce receiver-relative packet scores, which drive selection, LDPC rate allocation, and skipping. The receiver combines prediction with delivered tiles and applies an explicit fallback when no codeword arrives. 
}
\label{fig:system}
\end{figure*}

The proposed \method{} framework addresses task-oriented video transmission over a bandwidth-constrained and unreliable wireless link. As illustrated in Fig.~\ref{fig:system}, the system comprises a visual feature encoder, a receiver-aware transmission controller, a wireless channel, and a stateful perception receiver. Rather than transmitting every frame or feature uniformly, \method{} selectively communicates information that is expected to improve the receiver's current perception.

At the sender, each video frame is encoded into spatial feature packets. A synchronized receiver twin maintains a local replica of the receiver's feature cache and task memory based on previous transmission outcomes. A shared world model predicts the evolution of the receiver's cached features, while sender-side verification identifies regions where prediction is preferable to retaining previous information. The receiver twin then estimates the task-level impact of refreshing different regions, providing packet importance scores for transmission decisions. Given the available channel budget and estimated channel reliability, the controller jointly determines packet selection and unequal error protection.

At the receiver, successfully decoded packets update the predicted feature cache, while missing packets are handled through prediction or predefined fallback rules. The updated features are processed by a stateful task head to produce perception outputs. Codeword acknowledgments enable the sender to reproduce receiver-state updates and maintain synchronization across frames. This closed-loop architecture integrates temporal prediction, receiver-relative packet prioritization, and channel-aware resource allocation to support robust perception under limited communication resources.

The detailed system components and problem formulation are presented as follows. Table~\ref{tab:notations} summarizes the principal notation used in Sections~\ref{sec:system} and~\ref{sec:method}.

\begin{table*}[!t]
\centering
\caption{Summary of principal notations.}
\label{tab:notations}
\footnotesize
\setlength{\tabcolsep}{4pt}
\renewcommand{\arraystretch}{1.13}
\begin{tabular}{@{}lp{0.37\textwidth}lp{0.37\textwidth}@{}}
\hline
\textbf{Symbol} & \textbf{Meaning} & \textbf{Symbol} & \textbf{Meaning} \\
\hline
$t,n,p$ & Frame, feature-cell, and packet indices & $N,D,P$ & Numbers of cells, feature dimensions, and packets \\
$G_h,G_w$ & Spatial feature-grid dimensions & $q,L=q/6$ & Payload bits per feature dimension; nominal codewords per packet \\
$z_t,c_t$ & Current source features and receiver feature cache & $a_t,h_t,\mathcal H_t$ & Cell ages, task memory, and stored receiver history \\
$S_t,\hat S_t$ & Receiver state and synchronized sender-side twin & $\pi(n),r_{p,t}$ & Cell-to-packet mapping and packet-delivery indicator \\
$\tilde c_t,\bar c_t$ & Receiver-available prediction and affine-warped cache & $c_t^{\rm wm},c_t^{\rm pred}$ & World-model candidate and sender-verified prediction \\
$d_t^{\rm aff},d_t$ & Affine and learned residual displacement fields & $\ell_t,b_{p,t}$ & Learned gate logits and tile-verification bits \\
$H,W$ & Number of history slots and transformer width & $\alpha_s,\nu_s,\eta_s$ & Normalized age, freshness, and history-validity inputs \\
$Y_t,Y_t^\star$ & Predicted and ground-truth task outputs & $Y_t^{\rm all},Y_t^{\rm none}$ & All-refresh and prediction-only counterfactual outputs \\
$v_{p,t},\epsilon_j$ & Packet task-impact score and codeword failure probability & $j_p,\mathcal J,u_j$ & Packet rate choice, available choices, and codeword-unit cost \\
$\bar\gamma,|h|^2$ & Average linear SNR and channel power gain & $U,U_{\rm ref},U_{\rm pay},U_t$ & Per-frame budget, reference, payload allowance, and actual cost \\
$\beta,\lambda$ & Normalized budget and Lagrange multiplier & $\mathcal L_{\rm WM},V$ & World-model training loss and source-feature variance scale \\
$\mu,T,Q$ & Transmission policy, evaluation horizon, and task-quality measure & $R,\warp$ & Affine-coefficient clipping bound and feature-warp operator \\
\hline
\end{tabular}
\end{table*}

\subsection{Feature Packets and Receiver State}
At time $t$, a frozen DINOv2 ViT-S/14 encoder~\cite{oquab2023dinov2} maps a camera frame to $N=G_hG_w$ patch features. Principal component analysis (PCA) reduces each feature to $D=128$ dimensions. A common 6-bit uniform quantizer defines the feature representation; \method{} discards the three least significant bits and reconstructs at the coarse-bin center. We write $z_t\in\mathbb R^{N\times D}$ for these dequantized, transmission-precision features. A packet contains one $2\times2$ tile, yielding $P=N/4$ packets and $4Dq=1536$ payload bits per packet for $q=3$.
The $2\times2$ granularity matches two 1536-bit feature tiles to one 3072-bit codeword at the 3-bit precision used here, while retaining spatially selective updates.

The receiver state $S_t=(c_t,a_t,h_t,\mathcal H_t)$ contains its feature cache, cell ages, task memory, and the history needed for prediction. Let $\pi(n)$ identify the packet containing cell $n$, and let $r_{p,t}\in\{0,1\}$ indicate successful delivery of packet $p$ (one if decoded, zero otherwise). With the prediction $\tilde c_t$ available at the receiver, the cache update is
\begin{equation}
c_t[n]=\begin{cases}
z_t[n],&r_{\pi(n),t}=1,\\
\tilde c_t[n],&r_{\pi(n),t}=0.
\end{cases}\label{eq:cache}
\end{equation}
Here $\tilde c_t=c_t^{\rm pred}$ when the current verification header arrives; otherwise it is the deterministic fallback in Section~\ref{sec:fallback}. The deterministic task head returns $Y_t=f(c_t,h_{t-1})$ and updates its memory. Received cells have age zero; other cells retain their delivery history with incremented age. For the primary task, semi-supervised video object segmentation, the first-frame masks specify up to $K=10$ objects and $Y_t\in[0,1]^{N\times(K+1)}$ includes background probabilities. We use a fixed, training-free label-propagation head~\cite{jabri2020spacetime,caron2021emerging}; its settings are given in Section~\ref{sec:setup}.

\subsection{Codeword Channel and Budget}
The link model uses quadrature phase-shift keying (QPSK) and 5G new radio (NR) LDPC codes~\cite{3gpp38212}. Each codeword contains 3072 payload bits and a 24-bit cyclic redundancy check (CRC). Rates $\{1/3,1/2,2/3,3/4\}$ correspond to coded lengths $\{9288,6192,4644,4128\}$ bits. One channel unit represents 516 coded bits, giving codeword costs $u_j\in\{18,12,9,8\}$ for the respective code rates listed above. AWGN codeword error rates are calibrated by Monte Carlo simulation in Sionna~\cite{hoydis2022sionna} with a 20-iteration belief-propagation decoder. The video experiments use the calibrated rates in a codeword-level success/erasure simulation.

For independent Rayleigh block fading, $h$ denotes the complex channel coefficient (distinct from the task-memory variable $h_t$), and $|h|^2\sim\mathrm{Exp}(1)$ is constant within each codeword and changes between codewords. SNR is defined per modulation symbol as $E_s/N_0$. With average SNR $\bar\gamma$ in linear units, the failure probability is
\begin{equation}
\epsilon_j(\bar\gamma)=\E_{|h|^2}\!\left[\epsilon_j^{\rm AWGN}(\bar\gamma|h|^2)\right].\label{eq:bler}
\end{equation}
The sender uses average-SNR error rates and has no instantaneous channel knowledge. There is no retransmission within the 100-ms frame interval. At $-5$, 0, and 10~dB average SNR, the rate-1/3 code fails with probabilities of approximately 0.90, 0.52, and 0.07, respectively.

The full digital reference budget is $U_{\rm ref}=9P$ units: every \emph{6-bit} feature packet at rate 2/3. This equals $2322P$ QPSK symbols per frame. A normalized budget $\beta$ permits $U=\operatorname{round}(\beta U_{\rm ref})$ units per frame, with rounding to the nearest integer. We evaluate $\beta=0.0413$ and $0.125$. All bandwidth percentages are normalized to this digital reference. 
Side-information conventions are specified in
Section~\ref{sec:verification}.

\subsection{Receiver-Relative Transmission Objective}
For a transmission policy $\mu$, ground-truth task output $Y_t^\star$, and task-quality metric $Q$, the desired objective is
\begin{equation}
\begin{aligned}
\max_{\mu}\quad&\frac{1}{T}\sum_{t=1}^{T}\E[Q(Y_t,Y_t^\star)]\\
\text{s.t.}\quad&\frac{1}{T}\sum_{t=1}^{T}\E[U_t]\leq U.
\end{aligned}\label{eq:problem}
\end{equation}
The expectation is over channel outcomes, and
$t=1,\ldots,T$ indexes frames after initialization at $t=0$. To make transmission decisions without current task labels, \method{} uses an immediate task-output proxy and enforces a per-frame budget.

\section{World-Model-Guided Packet Transmission}
\label{sec:method}

Building on the system model and transmission objective established in Section~\ref{sec:system}, this section presents the technical implementation of \method{}. The following subsections detail receiver-relative task-impact estimation, world-model architecture and training, sender-side verification, packet selection and protection, and receiver operation under packet losses.

\subsection{Receiver Twin and Task-Impact Scores}\label{sec:twin}
The sender uses codeword decoding outcomes fed back
by the receiver to synchronize the receiver twin.
The twin evaluates two current-frame counterfactuals using the same task memory: $Y_t^{\rm all}=f(z_t,h_{t-1})$ with all cells refreshed, and $Y_t^{\rm none}=f(c_t^{\rm pred},h_{t-1})$ with no refreshed cells. The prediction-only counterfactual assumes the current verification information is available, while complete frame outages are handled separately in Section~\ref{sec:fallback}. Packet $p$ receives score
\begin{equation}
v_{p,t}=\sum_{n:\pi(n)=p}\sum_{k=1}^{K}
\left|Y_t^{\rm all}[n,k]-Y_t^{\rm none}[n,k]\right|.\label{eq:value}
\end{equation}
Background is excluded because its probability is determined by the object probabilities. Two task evaluations produce scores for all $P$ packets. The score spatially attributes the all-versus-none output discrepancy to packets, providing an efficient proxy for immediate task impact. Regions already reproduced by the predicted receiver state receive low priority. This includes predictable background motion and already tracked objects, which may still have large feature changes or high sender confidence.

\subsection{World-Model Architecture}\label{sec:wm}
\begin{figure*}[!tb]
\centering
\includegraphics[width=0.98\textwidth]{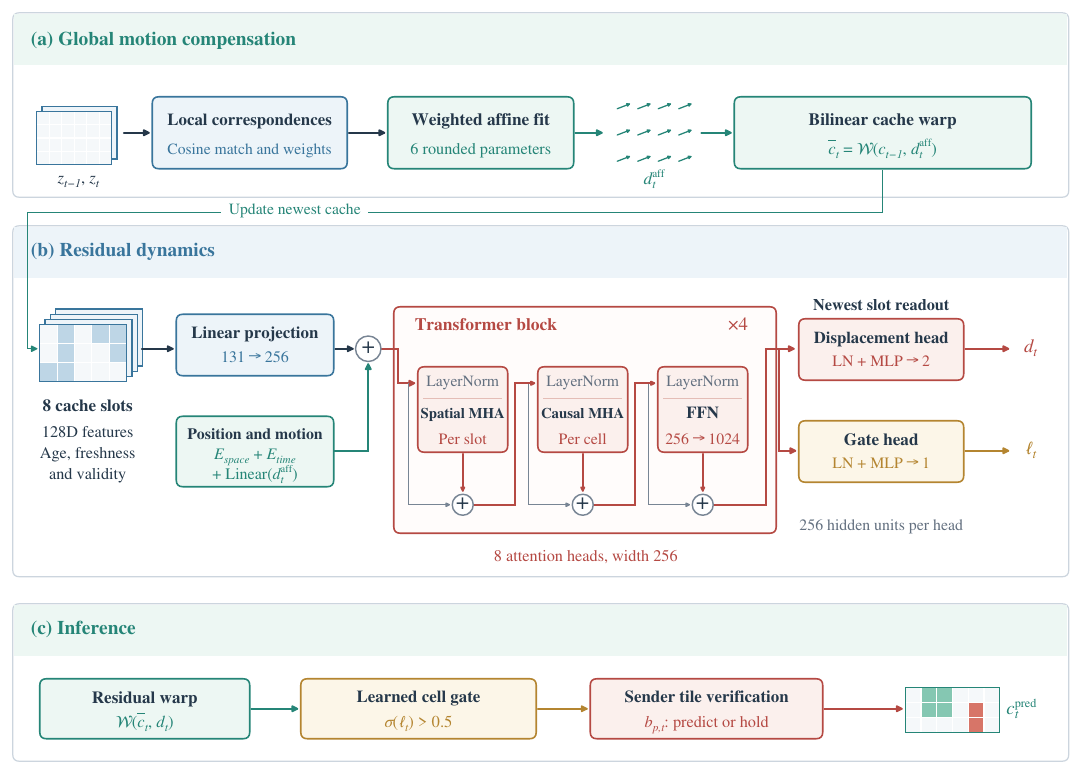}
\caption{Architecture of the proposed world model. (A) Sender-estimated affine motion advances the cached features. (B) Eight receiver-history slots are embedded with age, freshness, validity, position, and affine-motion information. Four factorized blocks apply spatial attention within each slot, causal temporal attention at each cell, and a feed-forward network; each sublayer uses pre-normalization and a residual connection. The newest-slot features feed displacement and gate heads. (C) Residual warping, learned cell gating, and sender-side tile verification are distinct operations. Training uses a soft gate, whereas inference applies a hard gating decision.}
\label{fig:wm_arch}
\end{figure*}

The world model predicts the evolution of cached visual features through motion. Fig.~\ref{fig:wm_arch} details its two stages. An affine field captures global camera motion; a learned residual model accounts for local displacement and decides whether warping is preferable to holding the cache.

\subsubsection{Affine motion compensation}
The sender obtains local correspondences between $z_{t-1}$ and $z_t$ using cosine similarity and a top-10 soft-argmax. Confidence-weighted least squares fits a six-parameter affine displacement field $d_t^{\rm aff}\in\mathbb R^{N\times2}$. Each affine coefficient is clipped to its prescribed symmetric range $[-R,R]$, where $R$ denotes the corresponding coefficient bound, and rounded with step $2R/255$. Given these parameters, both ends compute
\begin{equation}
\bar c_t=\warp(c_{t-1},d_t^{\rm aff}),\label{eq:affine}
\end{equation}
where $\warp(c,d)[n]$ samples $c$ at the two-dimensional grid coordinate $\boldsymbol p_n+d[n]$, using bilinear interpolation and replicated borders. Here $\boldsymbol p_n$ is the row-column coordinate of cell $n$; displacement maps a destination cell to its source location. The sender estimates the motion using its consecutive observations, while the receiver applies it to its possibly stale cache. The resulting field aligns the cached scene with the current global motion.
The receiver reconstructs the affine field from the same quantized coefficients carried in the decoded side information, rather than independently estimating motion.

\subsubsection{History embedding and spatiotemporal backbone}
The residual model uses $H=8$ history slots, distinct from the receiver-history variable $\mathcal H_t$. At each cell, a token concatenates the 128-dimensional cached feature, age clipped at 16 and divided by 16, a freshness flag, and a history-validity flag. A linear layer maps these 131 values to width $W=256$. Learned spatial and temporal embeddings are added, together with a projected two-dimensional affine field broadcast across history slots:
\begin{equation}
\begin{split}
x_{s,n}^{(0)}={}&W_{\rm in}[\tilde c_s^{\rm hist}[n];\alpha_s[n];\nu_s[n];\eta_s]\\
&+e_n^{\rm sp}+e_s^{\rm tm}+W_{\rm ego}d_t^{\rm aff}[n]+b_{\rm emb}.
\end{split}\label{eq:embedding}
\end{equation}
Here $s=1,\ldots,H$ indexes window positions, $\alpha_s[n]=\min(a_s[n],16)/16$, $\nu_s[n]=\mathbf1[a_s[n]=0]$, and $\eta_s$ indicates an available history slot. The input $\tilde c_s^{\rm hist}$ equals the stored cache except in the newest slot, which is replaced by $\bar c_t$; ages are unchanged. The vector $b_{\rm emb}$ combines the two linear-layer biases. At the start of a clip, the earliest available history is repeated to fill the window; validity is provided as an input flag rather than an attention-padding mask.

Four transformer blocks process these tokens. Each block applies spatial self-attention across the $N$ cells of each history slot, causal temporal attention across the $H$ slots of each cell, and a feed-forward network of widths $256\rightarrow1024\rightarrow256$. All attention modules use eight heads, pre-layer normalization, and residual connections. Factorizing attention gives an attention cost proportional to $HN^2+NH^2$, instead of $(HN)^2$ for joint attention over the full space-time grid.

\subsubsection{Displacement and gate heads}
The newest-slot features feed two separate heads, each comprising layer normalization, a 256-unit linear layer, GELU, and an output layer. They predict a residual displacement $d_t[n]\in\mathbb R^2$ and a gate logit $\ell_t[n]$. The inference-time candidate is
\begin{equation}
c_t^{\rm wm}[n]=
\begin{cases}
\warp(\bar c_t,d_t)[n],&\sigma(\ell_t[n])>1/2,\\
c_{t-1}[n],&\text{otherwise}.
\end{cases}\label{eq:wm_gate}
\end{equation}
The gate chooses between residual-warped features and the original, unwarped cache. 

\subsection{Label-Free Training of the World Model}\label{sec:wm_train}
Training uses simulated receiver histories from unlabeled video, rather than only clean consecutive frames. In this subsection, $z_t$ denotes the dequantized 6-bit source features used for world-model training; deployment uses the coarsened 3-bit payload described above. For each clip, the packet-loss probability is zero with probability 0.1 and otherwise uniform on $[0,0.95]$. In 87.5\% of training samples, the history ends 2-8 frames before the target, exposing the model to stale observations. These simulated histories include missing packets and affine motion compensation.

During training, a soft gate permits end-to-end differentiation:
\begin{equation}
c_t^{\rm soft}=\sigma(\ell_t)\odot\warp(\bar c_t,d_t)
+[1-\sigma(\ell_t)]\odot c_{t-1}.\label{eq:softgate}
\end{equation}
Let $V=\sum_{d=1}^{D}\operatorname{Var}(z[d])$ be the total feature variance, with variance computed over source features in the training data and $z[d]$ denoting feature dimension $d$. The prediction term is
\begin{equation}
\mathcal L_{\rm pred}=\frac{1}{N}\sum_n
\frac{\|c_t^{\rm soft}[n]-z_t[n]\|_2^2}{V}.\label{eq:predloss}
\end{equation}
A correspondence teacher supplies residual-displacement targets after affine alignment. Matches with confidence below 0.3 or forward-backward error above one grid cell are filtered. The term $\mathcal L_{\rm Huber}$ sums the unit-threshold smooth-$L_1$ loss over the two displacement coordinates and averages over retained matches (zero if none remain). A second prediction step uses the model's first output at missing cells and simulated received features at refreshed cells. The complete objective is
\begin{equation}
\mathcal L_{\rm WM}=\mathcal L_{\rm pred}
+0.1\mathcal L_{\rm Huber}+\mathcal L_{\rm rollout}.\label{eq:wmloss}
\end{equation}
The rollout term uses the same variance-normalized feature error at the next step; training averages the objective over sampled windows. The model is initialized from a displacement model trained with correspondence supervision and fine-tuned for ten epochs using AdamW with nominal learning rate $2\times10^{-4}$, a 500-step linear warmup multiplied by cosine decay, and weight decay 0.05. The world model is trained on unlabeled video sequences disjoint from evaluation data, with separate models trained for each domain (e.g., autonomous driving and UAV-based aerial perception).

\subsection{Sender Verification and Side Information}\label{sec:verification}
The sender can compare the candidate against its current features. For tile $p$, let $e_{p,t}^{\rm wm}$ and $e_{p,t}^{\rm hold}$ be the summed squared feature errors of $c_t^{\rm wm}$ and $c_{t-1}$ relative to $z_t$. It sends
\begin{equation}
\begin{aligned}
b_{p,t}&=\mathbf1[e_{p,t}^{\rm wm}<e_{p,t}^{\rm hold}],\\
c_t^{\rm pred}[n]&=
\begin{cases}
c_t^{\rm wm}[n],&b_{\pi(n),t}=1,\\
c_{t-1}[n],&b_{\pi(n),t}=0.
\end{cases}
\end{aligned}\label{eq:verify}
\end{equation}
Verification guarantees no larger \emph{tile feature error} than holding, when its side information arrives. The twin then evaluates task-output changes relative to this verified prediction, connecting feature-level verification to task-level packet selection.

The modeled side channel repeats the $P$-bit mask and affine parameters with every codeword, making them available whenever any codeword decodes. For $L=q/6$, the evaluation reserves a mask share in scaled allocation units:
\begin{equation}
U_{\rm pay}=L\left\lfloor\frac{U}{L}\left(1-\frac{P}{3072}\right)\right\rfloor.
\label{eq:paybudget}
\end{equation}
Whole-codeword packing then enforces $U$, as specified below. This codeword-level evaluation uses a budget-share approximation for the verification mask; affine coefficients, packet identifiers, and ACK traffic lie outside the reported forward budget. Side-information availability is coupled to codeword decoding.

\subsection{Packet Selection, UEP, and Packing}\label{sec:uep}
Each packet chooses $j_p\in\mathcal J=\{\mathrm{skip},3/4,2/3,1/2,1/3\}$. Skip has zero cost ($u_{\mathrm{skip}}=0$) and failure probability one ($\epsilon_{\mathrm{skip}}=1$). A packet with $q$ bits per feature dimension occupies $L=q/6$ nominal codewords relative to the 6-bit, 3072-payload-bit codeword (thus $L=1/2$ when $q=3$). Given the payload budget $U_{\rm pay}$, packet allocation minimizes the additive proxy
\begin{equation}
\begin{aligned}
\min_{\{j_p\}}\quad&\sum_{p=1}^{P}v_{p,t}\epsilon_{j_p}(\bar\gamma)\\
\text{s.t.}\quad&\sum_{p=1}^{P}L u_{j_p}\leq U_{\rm pay}.
\end{aligned}\label{eq:mck}
\end{equation}
This multiple-choice knapsack objective treats the scores as additive. Under Lagrangian relaxation, packet $p$ selects
\begin{equation}
j_p(\lambda)=\arg\min_{j\in\mathcal J}
\{v_{p,t}\epsilon_j(\bar\gamma)+\lambda L u_j\}.\label{eq:lagrange}
\end{equation}
Bisection searches for a feasible multiplier, after which remaining units fund greedy upgrades ordered by expected loss reduction per unit. The implementation adds $10^{-6}$ to every score to avoid zero-valued allocation ties. Packets with the same rate are ordered by value and packed into whole codewords. Writing $n_j$ for the number assigned rate $j$, packing enforces the separate physical-codeword constraint
\begin{equation}
U_t=\sum_{j\in\mathcal J\setminus\{\mathrm{skip}\}}u_j\lceil L n_j\rceil\leq U.
\label{eq:packing}
\end{equation}
If whole-codeword rounding violates this constraint, low-value packets are removed and packing is recomputed, freeing complete codewords as necessary until it holds. Thus packed cost can exceed $U_{\rm pay}$ while remaining within $U$. At $q=3$, two feature packets share each codeword and therefore share its erasure outcome. Linearity of expectation preserves the additive proxy in~\eqref{eq:mck} under these shared erasures. Lagrangian relaxation and packing provide an efficient approximation for the resulting discrete allocation.

\subsection{Operation During a Frame Outage}\label{sec:fallback}
If at least one codeword arrives, the receiver obtains the current side information, constructs the verified prediction, and overwrites the successfully delivered tiles. If no codeword arrives, it has neither the current verification mask nor the current affine parameters. It reuses the last delivered affine field and suppresses the unverified residual correction. If no affine field has ever arrived in the clip, it initially holds the cache; from frame index five onward, it uses the residual model with a zero affine condition and its learned hard gate. These rules also apply at the receiver twin using the reported codeword decoding outcomes. The receiver therefore maintains a continuous state trajectory across intermittent frame losses.

\subsection{Payload Precision}\label{sec:precision}
A 3-bit payload doubles spatial coverage per feature codeword relative to a 6-bit payload, at the cost of reduced feature precision. The choice of payload precision is evaluated in Section~\ref{sec:setup}.



\section{Experimental Setup}\label{sec:setup}

\subsection{Data, Tasks, and Evaluation Protocol}

\emph{KITTI-MOTS (KITTI)}~\cite{voigtlaender2019mots} provides driving video with instance masks. Sequences are partitioned into non-overlapping 32-frame clips, corresponding to 3.2~s at 10 frames/s. We track up to ten largest first-frame objects, retaining objects of area at least 400 pixels. Frames are resized to $224\times896$, giving a $16\times64$ feature grid. Development uses 141 clips from 12 training sequences; final evaluation uses 80 clips from 9 validation sequences. The world model is trained on 10,975 unlabeled frames from the official test partition, without using evaluation-sequence frames.

\emph{VisDrone2019-MOT (VisDrone)}~\cite{zhu2022visdrone} tests the approach on UAV video. Every third frame is used, giving approximately 10 frames/s, with the same 32-step clip length. Frames are resized to $392\times672$ ($28\times48$ cells, $P=336$ packets). We retain up to ten first-frame objects with box area at least 1600 pixels. Since the annotations are boxes, the mask-based task uses filled boxes as proxy masks. Development uses the training partition, which also provides 7,950 unlabeled frames for world-model training; evaluation uses 68 clips from 17 test-dev sequences.

For KITTI, the full digital reference corresponds to 5.94~Msymbol/s at 10 frames/s, and the evaluated budgets provide 95 and 288 channel units per frame. The verification-mask reservation accounts for 8.3\% of the nominal budget on KITTI and 10.9\% on VisDrone, before rounding.

The fixed label-propagation head compares features by cosine similarity at temperature 0.03, takes the top five matches within a radius of 16 grid cells on KITTI or 6 on VisDrone, and retains the first frame and the latest three memory frames. Settings are shared by communication methods within each dataset. The first-frame object labels are supplied for all methods. 

Reported comparisons use fixed configurations and matched evaluation clips, with one channel realization per clip and operating point. Random seeds are indexed by clip and SNR; each method applies its own codeword packing or analog noise generation. Precision and model settings are selected during development, and final scores are reported on the designated validation or test-dev partition.

\subsection{Metrics and Comparators}

Segmentation quality is the clip-averaged J\&F score~\cite{perazzi2016benchmark,pont2017davis}. Our evaluator averages region intersection-over-union and boundary F-score, first over interior frames and tracked objects within each clip, then equally over clips. The first and last frames are excluded from scoring. Feature-grid probabilities are bilinearly upsampled before taking the most likely label. Boundary matching uses tolerance $\lceil0.008\sqrt{H_e^2+W_e^2}\rceil$ pixels at the evaluation resolution $H_e\times W_e$; VisDrone masks wider than 960 pixels are resized to width 960 with aspect ratio preserved. All differences are reported in percentage points (pp) and are computed from the rounded scores given in the tables and figures. Rayleigh experiments span $-5$ to 20~dB average SNR; AWGN experiments span $-2$ to 4~dB, including the decoding transition. Paired bootstrap intervals~\cite{efron1993bootstrap} resample clips from each source sequence, holding the selected comparator fixed.

The independent baselines are as follows.
\begin{itemize}
\item \emph{H.264/H.265 + LDPC}: libx264/libx265 with zero-latency settings and no B-frames, operating at the receiver's input resolution. Network abstraction layer (NAL) units are packetized into 384-byte payloads and fail if any constituent codeword is lost; FFmpeg conceals missing content. Candidate configurations vary refresh policy, channel rate, and, for H.265, rate-control fill. We report the highest mean J\&F among configurations using no more than 105\% of the nominal budget, selecting the best admissible codec configuration at each evaluation operating point. Decoded frames pass through the same feature extractor and task head.

\item \emph{DeepJSCC}: a feature-tile JSCC baseline based on the learned joint source-channel coding principle in~\cite{bourtsoulatze2019deep}. Two 1024-unit hidden layers encode and decode each tile; the decoder also receives SNR and channel gain. Label-free feature mean-squared error (MSE) training uses 6000 AdamW steps at learning rate $10^{-3}$, with SNR sampled from $[-2,10]$~dB for AWGN or $[-5,25]$~dB for Rayleigh fading. Every tile uses 96 or 290 complex symbols at the two budgets.

\item \emph{Feature change} and \emph{sender confidence}: packet priorities use cache-feature differences or sender object confidence, respectively. Both use the digital rate allocator with their own priorities and a hold-last receiver.

\item \emph{Frame trigger + WM}: a twin-based frame discrepancy triggers whole-frame transmission with accumulated budget credit. The best evaluated threshold is reported. The baseline uses the displacement predictor and ideal verification availability.

\item \emph{Cyclic}, \emph{periodic}, and \emph{random}: packets are refreshed in round-robin order, complete frames are sent when sufficient credit accumulates, or packet priorities are randomized.
\end{itemize}

\emph{EEP at 100\%} sends every 3-bit packet at rate 1/3. Its budget equals the fixed 6-bit/rate-2/3 reference and serves as the full-budget reference.

All packet-selection baselines use the same 3-bit payload as \method{} on both datasets.

Payload precision is evaluated on 16 training-split clips using equal error protection at 4-dB AWGN and the full reference budget. The calibrated codeword error rate of the rate-1/3 code used here is below $10^{-3}$ at this operating point. As shown in Table~\ref{tab:precision}, the 6-bit and 3-bit representations achieve 55.68\% and 54.44\% J\&F, respectively, a reduction of 1.24 percentage points; the 2-bit representation achieves 52.97\%. These results support the 3-bit operating point as a favorable trade-off between feature precision and spatial coverage.

\begin{table}[!t]
\centering
\caption{Payload-precision sensitivity. Segmentation: 16 training clips, EEP under 4-dB AWGN, reference budget 100\%. Detection: clean validation features.}
\label{tab:precision}
\small
\begin{tabular}{@{}lrrrr@{}}
\toprule
Precision (bits/dimension)&6&4&3&2\\
\midrule
Segmentation J\&F (\%)&55.68&54.68&54.44&52.97\\
Detection mAP@0.5 (\%)&64.68&64.27&63.53&61.28\\
\bottomrule
\end{tabular}
\end{table}

\section{Results and Analysis}\label{sec:results}

\subsection{KITTI Results}\label{sec:kitti_results}

\subsubsection{Rayleigh Fading}

\begin{figure*}[!tb]
\centering
\includegraphics[width=0.98\textwidth]{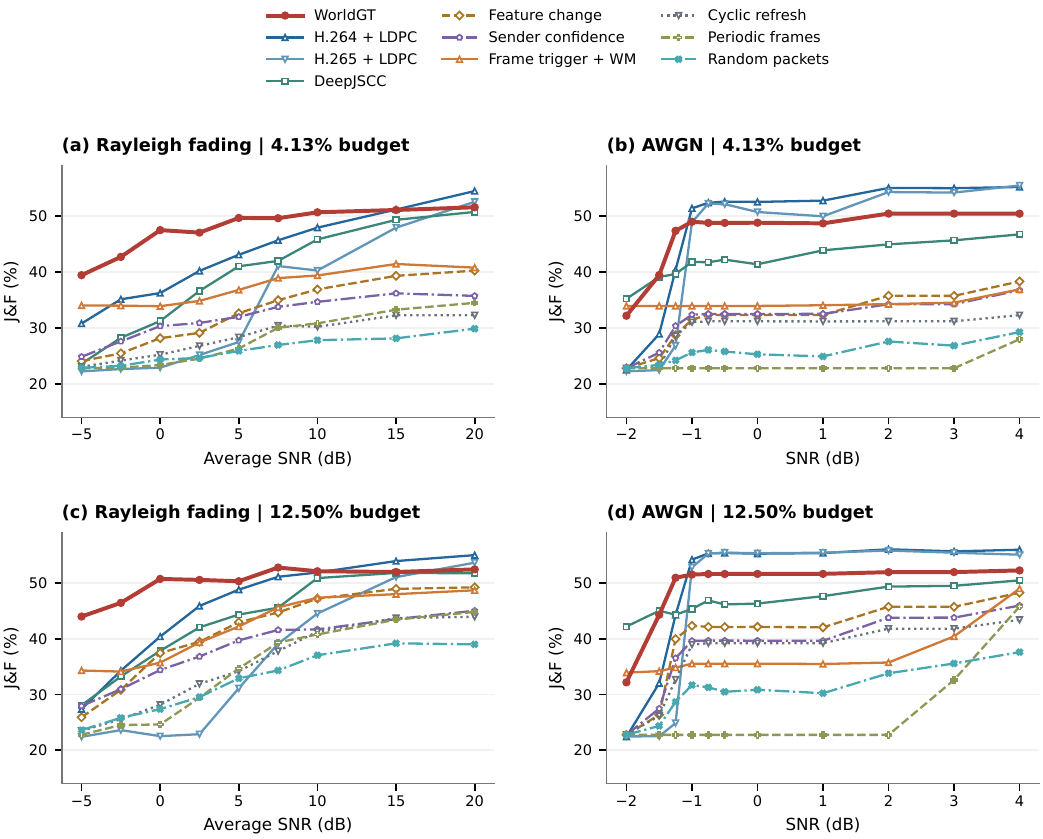}
\caption{KITTI segmentation compared with all nine baselines under Rayleigh fading (left) and AWGN (right), at 4.13\% (top) and 12.5\% (bottom) reference budgets. Each curve represents one method. WorldGT and the six packet-selection baselines use 3-bit feature payloads; the video codecs transmit encoded frames and DeepJSCC transmits analog features. 
}
\label{fig:snr_kitti}
\end{figure*}

Fig.~\ref{fig:snr_kitti} compares KITTI segmentation quality against all nine coding-based transmission and packet-selection baselines, each plotted individually. The four panels show Rayleigh fading and AWGN at both budgets. Under Rayleigh fading, \method{} outperforms all evaluated baselines at every measured SNR from $-5$ through 10~dB at both budgets. At 4.13\% budget, it achieves 39.5\% J\&F at $-5$~dB and 47.5\% at 0~dB, exceeding H.264 at these points by 8.7 and 11.3 percentage points, respectively. At 12.5\% budget and 0~dB, \method{} reaches 50.7\%, compared with 40.4\% for H.264 and 37.8\% for DeepJSCC. 

The low-SNR advantage is particularly relevant to intermittent delivery: even the most robust LDPC configuration loses approximately 90\% of its codewords at $-5$~dB and 52\% at 0~dB. The world model maintains a predictive receiver state between successful updates, while receiver-relative selection directs the available payload to features that improve the task output. Together, these mechanisms sustain useful segmentation despite frequent codeword losses.

At higher SNR, \method{} remains competitive as conventional codecs benefit from increasingly reliable delivery. At 20~dB, it reaches 51.6\% and 52.4\% J\&F at the two budgets, within 2.9 and 2.6 points of the best baseline, respectively. At 4.13\% budget, its 51.6\% score is also within 1.1 points of the full-budget 3-bit EEP reference (52.7\%). The resulting profile combines a substantial advantage under severe fading with competitive performance at high SNR.

The individual packet-selection curves in Fig.~\ref{fig:snr_kitti} further illustrate the advantage of targeted refresh. At 0~dB Rayleigh SNR and 4.13\% budget, \method{} achieves 47.5\% J\&F, compared with 33.9\% for frame triggering, 28.2\% for feature change, and 30.3\% for sender confidence. Cyclic, periodic, and random refresh remain below these task-adaptive approaches at this operating point. Section~\ref{sec:ablation} further isolates the contributions of selection and prediction at this budget.

\subsubsection{AWGN}

The AWGN panels of Fig.~\ref{fig:snr_kitti} (in the right column) show how prediction and targeted refresh support segmentation near the decoding transition. At $-1.25$~dB, \method{} achieves 47.4\% and 50.9\% J\&F at the two budgets, exceeding both video codecs and DeepJSCC. H.264 reaches 40.4\% and 44.3\% at the same operating points. At $-1.5$~dB, \method{} exceeds H.264 by 10.6 and 12.3 points, respectively. 

Temporal prediction also sustains the receiver during complete digital outage. At $-2$~dB, where all four calibrated LDPC options fail, \method{} achieves 32.2\% J\&F using initialization and predictive fallback, compared with 22.8\% for the matched hold-last receiver. This 9.4-point improvement isolates the value of advancing the cached scene when no new digital payload arrives. With reliable delivery at 4~dB, \method{} reaches 50.5\% and 52.2\% at the two budgets, within 5.0 and 3.7 points of the strongest codec. Its largest benefit is therefore concentrated around the difficult transition from outage to reliable packet reception.

\subsection{VisDrone Results}

\begin{figure*}[!tb]
\centering
\includegraphics[width=0.98\textwidth]{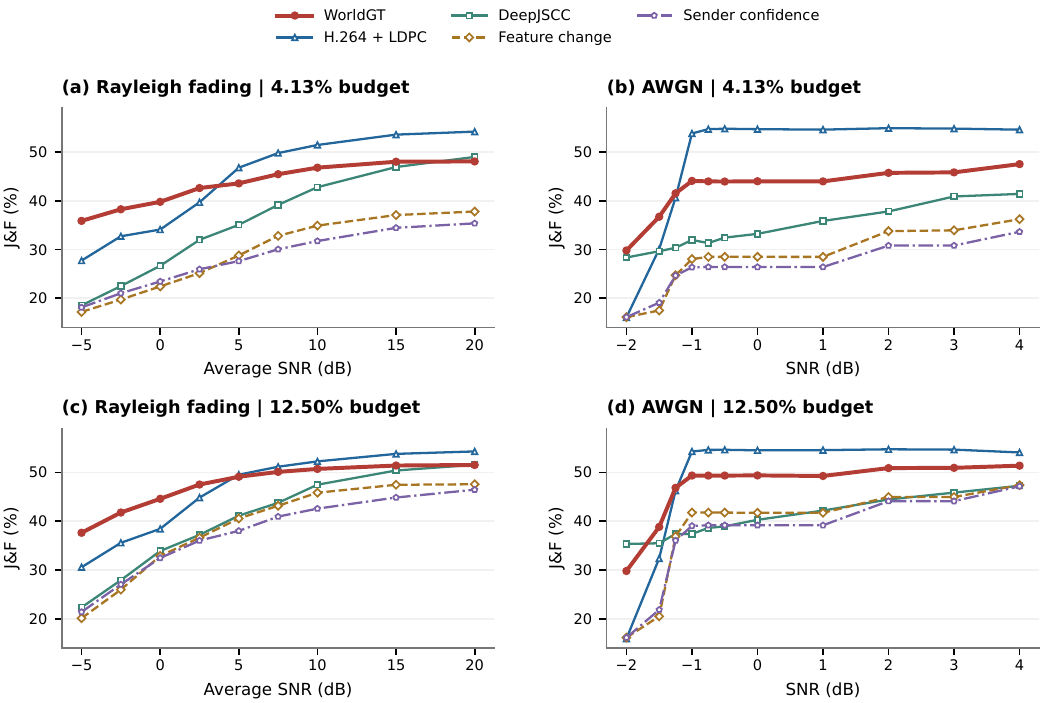}
\caption{VisDrone test-dev segmentation under Rayleigh fading (left) and AWGN (right), at 4.13\% (top) and 12.5\% (bottom) reference budgets. H.264 + LDPC, DeepJSCC, feature change, and sender confidence are plotted individually. J\&F uses filled bounding boxes as proxy masks. 
}
\label{fig:snr_visdrone}
\end{figure*}

Fig.~\ref{fig:snr_visdrone} compares segmentation quality on aerial video under both channel models and bandwidth budgets. Under Rayleigh fading, \method{} leads every evaluated baseline from $-5$ through 2.5~dB at both budgets. At $-5$~dB, it achieves 35.9\% and 37.6\% J\&F, exceeding H.264 by 8.2 and 7.0 points. At 0~dB, the scores increase to 39.8\% and 44.6\%, compared with 34.1\% and 38.4\% for H.264 and 26.7\% and 33.9\% for DeepJSCC. These results extend the low-SNR advantage from driving scenes to aerial video. Appendix 

The feature-change and sender-confidence baselines further demonstrate the value of receiver-relative selection. At 0~dB Rayleigh SNR and 4.13\% budget, they achieve 22.4\% and 23.5\% J\&F, respectively, compared with 39.8\% for \method{}. The comparison holds payload precision fixed, associating the improvement with the predictive receiver and its task-aware refresh decisions. At 12.5\% budget, \method{} remains within 0.4 points of H.264 at 5~dB and within 2.8 points at 20~dB.

Under AWGN, \method{} leads the evaluated baselines at $-1.5$ and $-1.25$~dB at both budgets. At $-1.5$~dB and 4.13\%, it reaches 36.7\% J\&F, compared with 30.1\% for H.264 and 29.7\% for DeepJSCC. At 12.5\%, it reaches 38.8\%, exceeding the same baselines by 6.4 and 3.3 points. Thus, the aerial-video experiments reproduce the benefit of predictive refresh near the decoding transition, alongside the broader low-SNR advantage under fading.

\subsection{Controlled Ablations and Mechanism Analysis}\label{sec:ablation}

\begin{figure*}[!tb]
\centering
\includegraphics[width=0.98\textwidth]{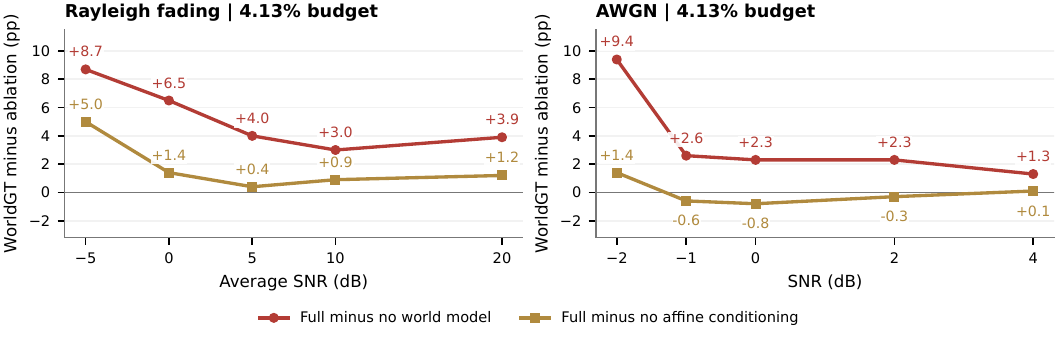}
\caption{Component contributions on KITTI at 4.13\% budget under Rayleigh fading (left) and AWGN (right). Each point subtracts the indicated ablation's clip-averaged J\&F from the complete \method{} score at the same SNR; annotations give the difference in percentage points. Positive values favor the complete method. Table~\ref{tab:ablation} gives absolute Rayleigh scores.}
\label{fig:ablation}
\end{figure*}

Fig.~\ref{fig:ablation} quantifies the contributions of the world model and affine conditioning under Rayleigh fading (left) and AWGN (right). For each channel and SNR, each plotted value is the complete method's clip-averaged J\&F minus that of the indicated ablation, evaluated on the same 80 KITTI clips at 4.13\% budget. The difference is expressed in percentage points; a positive value indicates the improvement supplied by the removed component. For example, at $-5$~dB Rayleigh SNR, the world-model contribution is $39.5-30.8=8.7$ points. Differences are computed from the rounded scores, consistent with Table~\ref{tab:ablation}. Table~\ref{tab:ablation} complements the figure with absolute Rayleigh J\&F scores and additional selection controls. The experiments fix the payload precision, channel model, task head, evaluation clips, and bandwidth budget.

\emph{World model.}
Replacing the predictive receiver with hold-last features reduces
Rayleigh J\&F by 8.7 points at $-5$~dB, 6.5 points at 0~dB,
and 3.0-4.0 points at 5-20~dB.
Prediction contributes both to receiver-cache updates and to
the reference used for packet selection.

\emph{Affine conditioning.}
Removing the affine stage and running the residual predictor without
that condition costs 5.0 points at $-5$~dB under Rayleigh fading.
The gain is smaller at higher Rayleigh SNRs.
This comparison captures the combined effects of affine alignment,
residual-model conditioning, and global-motion persistence during outages.

\emph{Selection without prediction.}
Without the world model, twin-based selection still reaches 41.0\%
at 0~dB Rayleigh SNR.
The frame-triggering baseline with the world model reaches 33.9\%
at the same operating point.
At 20~dB, the corresponding scores are 47.7\% and 40.8\%.
The complete method reaches 47.5\% and 51.6\% at these two
operating points, respectively.
Receiver-relative selection and temporal prediction therefore
provide complementary gains.

\emph{UEP and payload.}
A development experiment found at most a 0.8-point reduction
when the best single rate per frame replaced per-packet UEP.
These results identify prediction and receiver-relative selection
as the main contributors, while payload coarsening increases
the number of spatial updates carried by each codeword.

\begin{table}[!tb]
\centering
\caption{KITTI Rayleigh fading at 4.13\% budget: J\&F (\%) for ablations and selection controls. The EEP row provides the 100\% budget reference.}
\label{tab:ablation}
\footnotesize
\setlength{\tabcolsep}{4pt}

\begin{tabular}{@{}lrrrrr@{}}
\toprule
SNR (dB) & $-5$ & 0 & 5 & 10 & 20\\
\midrule

\rowcolor{worldwash}\method{} & 39.5 & 47.5 & 49.7 & 50.7 & 51.6 \\
Without affine conditioning & 34.5 & 46.1 & 49.3 & 49.8 & 50.4 \\
Without world model & 30.8 & 41.0 & 45.7 & 47.7 & 47.7 \\
Frame trigger + WM & 34.0 & 33.9 & 36.8 & 39.4 & 40.8 \\
Feature change & 24.1 & 28.2 & 32.6 & 36.9 & 40.3 \\
Sender confidence & 24.8 & 30.3 & 32.0 & 34.7 & 35.7 \\
Cyclic refresh & 23.0 & 25.2 & 28.3 & 30.2 & 32.3 \\
Periodic full frames & 22.8 & 23.4 & 26.3 & 30.8 & 34.5 \\
Random packets & 22.8 & 24.3 & 25.9 & 27.8 & 29.9 \\
\midrule

EEP (100\% budget) & 30.6 & 46.2 & 51.5 & 52.1 & 52.7 \\
\bottomrule
\end{tabular}
\end{table}

\subsection{Joint Bandwidth-SNR Trade-off}\label{sec:rate_snr}

\begin{figure*}[!tb]
\centering
\includegraphics[width=\textwidth]{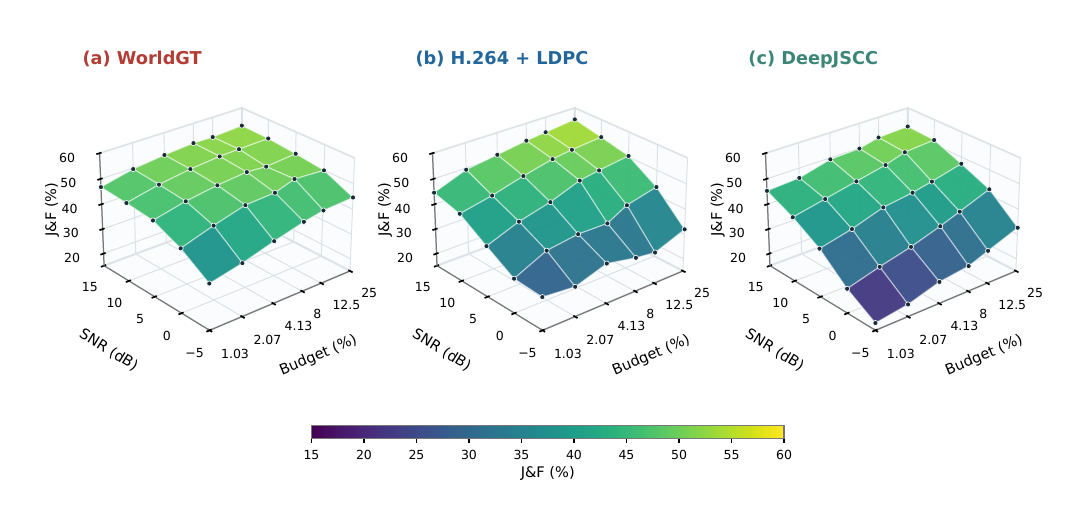}
\caption{J\&F-bandwidth-SNR surfaces on KITTI under Rayleigh fading, evaluated on 80 clips per operating point. Panels (a)-(c) share viewing angles, axis limits, and color scales; bandwidth is shown on a logarithmic coordinate. Dots mark the $5\times6$ measurement grid, with connecting facets for visualization.}
\label{fig:rate_snr}
\end{figure*}

Fig.~\ref{fig:rate_snr} compares the J\&F-bandwidth-SNR surfaces of \method{}, H.264, and DeepJSCC on the same 80 KITTI clips under Rayleigh fading. Each surface contains 30 measured operating points: SNRs of $-5$, 0, 5, 10, and 15~dB, combined with budgets of 1.03\%, 2.07\%, 4.13\%, 8\%, 12.5\%, and 25\%. The 4.13\% and 12.5\% points reuse the SNR sweeps; the additional budgets use the same method configurations, and H.264 follows the budget-admissible selection rule in Section~\ref{sec:setup}.

\method{} exceeds both competitors across the entire bandwidth range at $-5$ and 0~dB. Even at 1.03\% budget, it retains 33.1\% J\&F at $-5$~dB and 40.2\% at 0~dB. The latter is close to H.264's 40.4\% at 12.5\% budget, using approximately twelve times less nominal bandwidth. At 5~dB, \method{} reaches 47.9\% at 2.07\%, exceeding H.264's 47.5\% at 8\%. These operating points illustrate how prediction and targeted refresh convert a limited transmission budget into useful task updates.

At 10~dB, \method{} remains ahead of both competitors through 12.5\% budget and is within 0.6 points of H.264 at 25\%. At 15~dB, it is nearly equal to H.264 at 4.13\%, with a 1.0-2.5-point gap at larger budgets. Across the joint grid, the surfaces show the strongest advantage when channel reliability or bandwidth is scarce, together with competitive performance as transmission resources increase.

\subsection{Qualitative Comparison of Received Perception}
\label{sec:qualitative}

\begin{figure*}[!tb]
\centering
\includegraphics[width=0.98\textwidth]{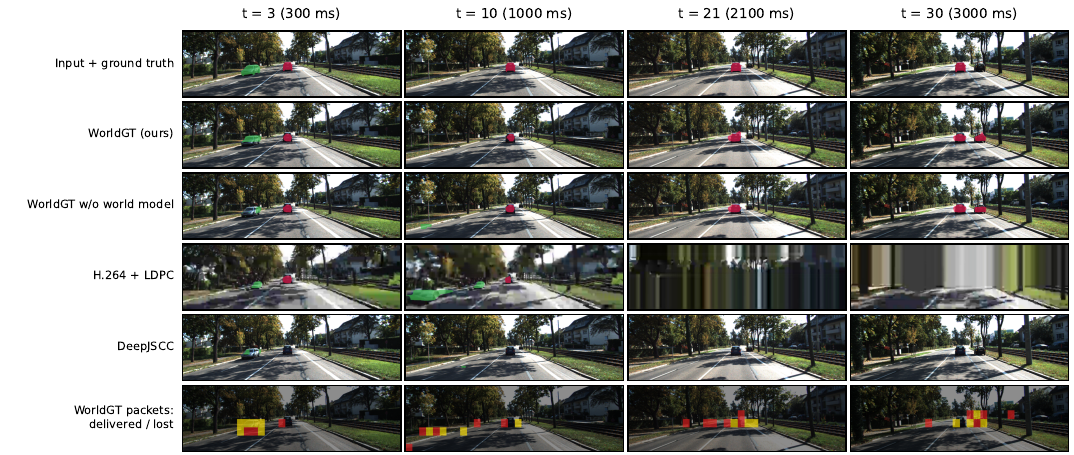}
\caption{Qualitative comparison on KITTI clip
\texttt{0010\_000128} at 0~dB average SNR and 4.13\%
reference budget. Columns show frames 3, 10, 21, and 30
after initialization. Colors identify object masks;
in the packet row, yellow denotes delivered tiles,
red denotes sent-but-lost tiles, and uncolored regions
were not sent. Feature-based predictions use the original
image as the visualization background, while H.264 masks
are overlaid on decoded frames. H.264 uses its selected
rate-1/3 IDR-refresh configuration.}
\label{fig:qualitative}
\end{figure*}

Fig.~\ref{fig:qualitative} presents a visual comparison on a
representative KITTI clip (\texttt{0010\_000128}) at 4.13\%
budget under 0~dB Rayleigh fading. The columns show four
frames at different time steps, allowing comparison of the
spatial consistency and temporal evolution of the predicted
object masks.

Across the displayed frames, \method{} maintains a localized
mask on the leading vehicle. Without the world model,
additional foreground regions appear around the vehicles,
particularly in the final frame. H.264 exhibits visible
decoding artifacts and severe frame corruption in later
frames, while DeepJSCC produces less accurate object masks
despite relatively clean visual backgrounds.
The final row illustrates the spatial distribution of
delivered, lost, and unsent packets, highlighting how
\method{} combines selective transmission with predicted
and cached information to maintain perception under
unreliable delivery.

\subsection{Computational Cost}

Table~\ref{tab:bench} reports the computational cost of
the profiled components on an NVIDIA GeForce RTX 4090
Laptop GPU. The microbenchmark uses one training clip
after warmup, with random packet scores, a 25\% reference
budget, and 3-dB AWGN error rates for allocation.
The sum of component median times is 93.2~ms on the
KITTI grid, with a peak memory usage of 0.15~GB
for the profiled components.
These measurements exclude feature extraction, the initial
affine warp, channel coding and decoding, transport,
and feedback, and therefore do not represent end-to-end
latency.


The Lagrangian allocator requires 30.9~ms, compared
with 544.8~ms for exact knapsack optimization,
corresponding to approximately $17.6\times$ faster
allocation. Exact knapsack optimization is used only
for comparison and is not part of the proposed
inference pipeline.

\begin{table}[!t]
\centering
\caption{Computational cost of profiled components on
the KITTI grid. Timing values are per-frame medians; the sum uses unrounded component medians.}
\label{tab:bench}
\small
\begin{tabular}{@{}lr@{}}
\toprule
Metric & Value\\
\midrule
Receiver task step & 6.6 ms\\
Two additional twin evaluations & 13.1 ms\\
Affine estimation & 6.5 ms\\
Residual predictor and warp & 34.7 ms\\
Tile verification & 1.3 ms\\
Lagrangian allocation & 30.9 ms\\
\midrule
Sum of reported component medians & 93.2 ms\\
\midrule
World-model parameters & 4.68 M\\
Peak memory of profiled components & 0.15 GB\\
\bottomrule
\end{tabular}
\end{table}

\subsection{Task Extension: Object Detection}
\label{sec:detection}

\begin{figure*}[!tb]
\centering
\includegraphics[width=0.98\textwidth]{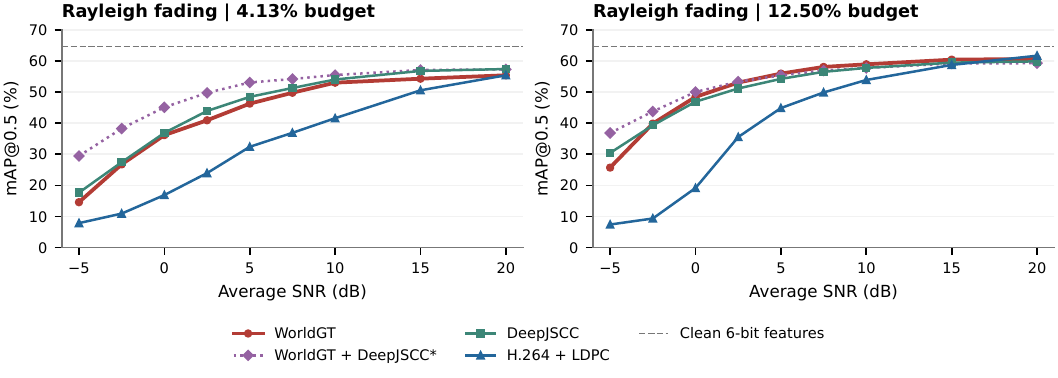}
\caption{Object detection on KITTI under Rayleigh fading at
4.13\% and 12.5\% bandwidth budgets. The detection head
is trained on clean features and applied directly to
features received through segmentation-oriented
communication policies. WorldGT + DeepJSCC combines
receiver-relative tile selection with analog feature
transmission. The dashed line indicates detection
performance with clean 6-bit features.}
\label{fig:detection}
\end{figure*}

Fig.~\ref{fig:detection} demonstrates the transferability
of WorldGT's segmentation-oriented communication strategy
to object detection across Rayleigh fading conditions
and two bandwidth budgets. A CenterNet-style detection
head~\cite{zhou2019objects} predicts car and pedestrian
heat maps, box sizes, and center offsets from the received
feature grid. The head has three layers with width 128
and is trained on clean features without subsequent
adaptation to the communication policies.
The scores and packet choices are inherited from the segmentation task, so the detection experiment evaluates transfer rather than detection-optimal selection.
Detection performance is measured by mAP at IoU 0.5
over both classes and all frames of the 80 KITTI
evaluation clips. The clean 6-bit feature reference
achieves 64.7\%.

At 12.5\% budget, \method{} outperforms
H.264 and all-tile DeepJSCC from $-2.5$ to 15~dB.
At 0~dB, it achieves 48.4\% mAP, exceeding DeepJSCC
by 1.5 points and H.264 by 29.2 points.
At 4.13\% budget, \method{} reaches 36.2\%
mAP at 0~dB, compared with 36.9\% for DeepJSCC
and 16.9\% for H.264.
At this budget, all-tile DeepJSCC retains a small
advantage, while \method{} remains well above
the digital H.264 baseline at low and moderate SNR.

We further extend WorldGT by integrating DeepJSCC
for analog feature transmission, yielding
\emph{WorldGT + DeepJSCC}.
This variant preserves the world model, receiver twin,
sender-side verification, and task-impact-based
packet selection, while replacing quantized feature
packets and LDPC coding with analog transmission.
It selects $B$ tiles according to receiver-relative
task-impact scores and allocates $k$ complex channel
symbols to each selected tile, while unselected
tiles are reconstructed using the verified predictor.
The normalized payload is $Bk/(2322P)$, with
$(B,k)=(\operatorname{round}(P/3),289)$ and
$(P/2,580)$ at the two budgets, respectively.
The sender twin uses the receiver's noisy reconstruction,
with verification available at both ends;
control traffic is excluded from the reported budget.

The resulting WorldGT + DeepJSCC achieves further
detection gains, demonstrating that WorldGT's
predictive selection mechanism is effective
across digital and analog transmission.
At 4.13\% budget, it outperforms DeepJSCC from $-5$ to 15~dB,
achieving 29.5\% mAP at $-5$~dB and 45.1\%
at 0~dB, corresponding to improvements of
11.8 and 8.2 points; at 20~dB the two are within 0.1 points.
At 12.5\% budget and 0~dB, it achieves
50.0\% mAP, compared with 46.9\% for DeepJSCC.
These results support the transferability of
receiver-relative predictive selection beyond
the original segmentation task and show its
potential as a general feature-transmission
strategy for downstream visual perception.

\section{Discussion}\label{sec:discussion}

The results demonstrate the value of making transmission
decisions relative to the receiver's predicted scene state.
Rather than treating each frame as an independent source
of information, \method{} exploits temporal predictability
to identify which spatial updates are most valuable for
downstream perception. Three design principles emerge.
First, packet utility should reflect its contribution
beyond what the receiver already knows, rather than
feature changes at the sender alone. Second, world-model
prediction and verification enable selective transmission
by distinguishing recoverable scene information from
updates that require communication. Third, payload
precision directly influences the spatial coverage
achievable under a constrained bandwidth budget.
Together, these mechanisms allow the receiver to maintain
task-relevant scene information despite substantial
packet losses.

The bandwidth-SNR experiments further reveal how the
relative importance of prediction and feature fidelity
changes with channel conditions. Under severe fading,
world-model prediction maintains useful scene estimates
between successful packet deliveries, while greater
channel reliability increases the value of transmitting
more detailed spatial information. The effectiveness of
the 3-bit operating point illustrates the benefit of
balancing feature precision against spatial coverage.
Beyond segmentation, the detection experiments show
that the transmitted feature representation supports
a separately trained detection head without retraining
the world model or redesigning the segmentation-oriented
transmission policy. The WorldGT + DeepJSCC results further demonstrate the
applicability of receiver-relative selection to analog
feature transmission. These findings open
opportunities for adaptive payload precision, hybrid
digital-analog communication, feedback-aware state
refresh, and longer-horizon packet selection. Extending
the world model to predict newly visible scene content
is another promising direction.

\section{Conclusion}\label{sec:conclusion}

This paper introduced \method{}, a world-model-guided
semantic packet transmission framework for video
perception over unreliable wireless channels.
By combining predictive receiver-state evolution,
sender-side verification, and receiver-relative
task-impact-based packet selection, \method{} focuses
communication resources on information that cannot
be adequately recovered from the receiver's existing
scene representation. A synchronized receiver twin
guides packet selection and unequal error protection,
enabling task-oriented transmission under stringent
bandwidth constraints.

Experiments on KITTI and VisDrone demonstrate substantial
perception gains under challenging channel conditions.
On KITTI under Rayleigh fading at a 4.13\% reference
budget, \method{} improves J\&F by 11.3 percentage
points over the strongest independent baseline at
0~dB and remains within 2.9 points at 20~dB.
Ablation studies confirm the contributions of
world-model prediction and receiver-relative selection,
while object-detection experiments demonstrate
transferability beyond the original segmentation
task. The results establish receiver-state-aware
predictive transmission as an effective approach
to bandwidth-efficient and reliable visual
semantic communication.

\FloatBarrier
\bibliographystyle{IEEEtran}
\bibliography{refs}

\end{document}